\documentclass[twocolumn]{aastex62}

\newcommand{\lbol}{L_{\rm bol}}

\usepackage{natbib}
\usepackage{amsmath}
\usepackage{multirow} 

\received{\today}
\submitjournal{ApJ}

\shorttitle{Drastic Luminosity Decline of dying AGN in Arp 187}
\shortauthors{Ichikawa et al.}

\begin{document}

\title{Discovery of Dying Active Galactic Nucleus in Arp~187: Experience of Drastic Luminosity Decline within $10^4$~years}

\correspondingauthor{Kohei Ichikawa}
\email{k.ichikawa@astr.tohoku.ac.jp}

\author[0000-0002-4377-903X]{Kohei Ichikawa}
\affil{Frontier Research Institute for Interdisciplinary Sciences, Tohoku University, Sendai 980-8578, Japan}
\affil{
Department of Astronomy, Columbia University, 550 West 120th Street, New York, NY 10027, USA}
\affil{
Department of Physics and Astronomy, University of Texas at San Antonio, One UTSA Circle, San Antonio, TX 78249, USA}

\author{Junko Ueda}
\affil{
National Astronomical Observatory of Japan, 2-21-1 Osawa, Mitaka, Tokyo 181-8588, Japan}
\affiliation{
Harvard-Smithsonian Center for Astrophysics, 60 Garden Street, Cambridge, MA 02138, USA}

\author{Hyun-Jin Bae}
\affil{
Department of Medicine, University of Ulsan College of Medicine, Seoul 05505, Republic of Korea
}

\author{Taiki Kawamuro}
\altaffiliation{JSPS fellow}
\affil{
National Astronomical Observatory of Japan, 2-21-1 Osawa, Mitaka, Tokyo 181-8588, Japan}

\author{Kenta Matsuoka}
\altaffiliation{JSPS fellow}
\affil{
Dipartimento di Fisica e Astronomia, Universit{\'a} degli Studi di Firenze, Via G. Sansone 1, I-50019 Sesto Fiorentino, Italy
}
\affil{
INAF -- Osservatorio Astrofisico di Arcetri, Largo Enrico Fermi 5, I-50125 Firenze, Italy
}
\affil{
Research Center for Space and Cosmic Evolution, Ehime University, 2-5 Bunkyo-cho, Matsuyama, Ehime 790-8577, Japan
}

\author{Yoshiki Toba}
\altaffiliation{JSPS fellow}
\affil{
Research Center for Space and Cosmic Evolution, Ehime University, 2-5 Bunkyo-cho, Matsuyama, Ehime 790-8577, Japan
}
\affil{
Academia Sinica Institute of Astronomy \& Astrophysics (ASIAA), 11F of Astronomy-Mathematics Building, AS/NTU, No.1, Section 4, Roosevelt Road, Taipei 10617, Taiwan
}
\affil{
Department of Astronomy, Kyoto University, Kitashirakawa-Oiwake-cho, Sakyo-ku, Kyoto 606-8502, Japan
}

\author{Megumi Shidatsu}
\affil{
Department of Physics, Faculty of Science, Ehime University, Matsuyama 790-8577, Japan
}



\begin{abstract}

Arp~187 is one of the fading active galactic nuclei (AGN),
whose AGN activity is currently decreasing in luminosity.
We investigate the observational signatures of AGN in Arp~187, which trace
 various physical scales from less than 0.1~pc to the nearly 10~kpc,
 to estimate the longterm luminosity change over $10^{4}$~years.
The VLA 5~GHz, 8~GHz, and the ALMA 133~GHz images 
reveal bimodal jet lobes with $\sim$5~kpc size and the absence of the central radio-core.
The 6dF optical spectrum shows 
that Arp~187 hosts  narrow line region with the estimated size of $\sim$1~kpc,
and the line strengths give the AGN luminosity 
of $L_{\rm bol}=1.5 \times 10^{46}$~erg~s$^{-1}$.  
On the other hand, the current AGN activity estimated 
from the AGN torus emission gives 
the upper bound of $L_{\rm bol} < 2.2 \times 10^{43}$~erg~s$^{-1}$.  
The absence of the radio-core gives the more strict upper bound 
of the current AGN luminosity of $L_{\rm bol} < 8.0 \times 10^{40}$~erg~s$^{-1}$, 
suggesting that the central engine is already quenched.
These multi-wavelength signatures indicate that 
Arp~187 hosts a ``dying'' AGN: the central engine is already dead, 
but the large scale AGN indicators are still observable 
as the remnant of the past AGN activity.  
The central engine has experienced the drastic luminosity decline 
by a factor of $\sim10^{3-5}$ fainter within $\sim10^{4}$~years, 
which is roughly consistent with the viscous timescale of 
the inner part of the accretion disk within $\sim$500~years.
 
\end{abstract}




\section{Introduction}\label{sec:intro}

One of the key questions in modern Astronomy is
 how supermassive black holes (SMBHs) and their host galaxies
co-evolve, leading to the tight correlation
between the masses of SMBHs and their bulges in the present universe \citep[e.g.,][]{kor13}.
Active galactic nuclei (AGN) are the best laboratories
 to understand this co-evolution process, since they are in the stage
 of the mass accretion onto SMBHs until the SMBHs reach their achievable maximum mass limit 
of $M_{\rm BH} \simeq 10^{10.5} M_{\odot}$ \citep{net03,mcl04,mcc11,kor13,tra14,jun15,wu15,ina16,ich17b}. 

One of the biggest unknown for this accretion process is 
 how long such AGN phase can last.
  Several authors indicate that 
 the total AGN phase has a duration of $10^{7-9}$ years \citep{mar04},
 and that one cycle of AGN should be at least 
 over $10^5$ years suggested from the observations \citep{sch15} and simulations \citep[e.g.,][]{nov11}.
This is also consistent with the results that there are various AGN indicators 
with different physical scales, and each AGN indicator has a tight luminosity correlations each other:
AGN nucleus (X-ray) and the 10~pc scale AGN dusty torus 
\citep[mid-infrared, hereafter MIR; ][]{gan09,ich12,asm15,mat15,ich17a}
and 1~kpc scale ionized gas region 
\citep[so-called narrow line region or NLR; e.g., ][]{net06,pan06,ued15,ber15}.
This long lifetime of AGN, however, implies a difficulty for us to 
observe the scene where AGN is now being quenched, or ``dying'',
which gives us the information how rapid the physics of accretion disk 
 in AGN is changing within the certain amount of time.

Recent observations, however, have discovered a key population of AGN to resolve the issue above.
While they show the AGN signatures at large physical scale with $\geq 1$~kpc (e.g., NLR and/or radio jets),
they lack the signatures at small physical scale with $<10$~pc (e.g., X-ray emission, the dust torus emission, and/or the radio core emission) 
or their luminosities are very faint even if they exist.
They are thought to be in the transient stage where their central engine 
has been already fading, but the large scale AGN indicators are still active
because of the long light crossing time (e.g., $>10^3$~years).
They are called fading AGN and $\sim20$ such sources
have been reported
\citep{sch10,sch13,sch16,sch13b,ich16,kee17,kaw17,vil18,wyl18,sar18a}.

Among them, Arp~187 located at $z=0.040$ ($D_{\rm L}$ = 178~Mpc, 1\arcsec = 798~pc) 
is one of the most promising ``dying'' AGN candidate, which is the final phase of fading AGN
whose current central engine is already quenched,
but the large scale AGN indicators are still alive because of the photon time delay.
\cite{ich16} used a jet lobe size discovered by 
the Atacama Large Millimeter/submillimeter Array (ALMA)
cycle-0 observation \citep{ued14} for estimating the upper limit of the 
quenching time of the fading AGN. 
Assuming a jet angle to the line of sight of 90$^\circ$ 
and a typical expansion, the kinematic age of the radio
 jets is estimated to be $8 \times 10^4$ year. 
 \cite{ich16} also revealed that AGN nucleus activity is
  already quenched with the bolometric
  luminosity of $L_{\rm bol} < 10^{41}$~erg~s$^{-1}$.
  However, \cite{ich16} could not estimate the past AGN luminosity,
  therefore we could not evaluate how rapid this AGN has experienced
  the luminosity decline.

In this paper, we report the more multi-wavelength support 
that Arp~187 hosts a bona fide dying AGN,
and the central engine has experienced drastic luminosity decline
over $10^{3-5}$ times within $10^4$~years, using the multiple-wavelength data 
including the newly obtained radio data with ALMA, 
the Karl G. Jansky Very Large Array (VLA), \textit{Spitzer},
and \textsc{Neowise} as well as the updated analysis method to the archival 6dF optical spectrum which has not been explored in our previous paper of \cite{ich16}.
Throughout the paper, we adopt $H_0 = 70.0$~km~s$^{-1}$~Mpc$^{-1}$,
$\Omega_{\rm M}=0.3$, and $\Omega_{\Lambda}=0.7$.

\section{Analysis and Results}\label{sec:analysis}

 
\begin{figure*}
\begin{center}
\includegraphics[width=0.95\textwidth]{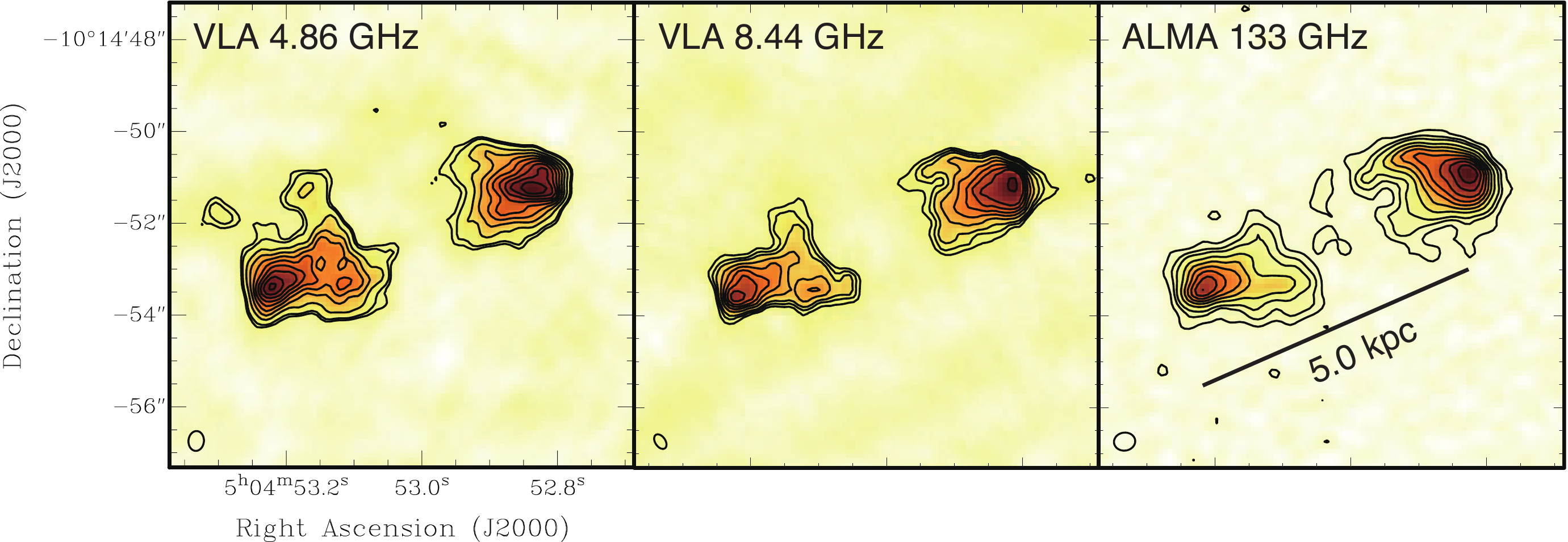}
\caption{
The radio continuum maps of Arp~187 
at VLA 4.86~GHz (left), VLA 8.44~GHz (middle), and ALMA 133~GHz (right).
The ellipse in the bottom-left corner shows the beam size measured in each map.
The center of each panel corresponds to the peak in the $K$-band image \citep{rot04}.
The length of each panel is 10\arcsec, which corresponds to $\sim$8~kpc.
The scale in the left panel shows the separation of the tow lobe cores ($\sim$5~kpc).
The contour levels are 
0.28~mJy~beam$^{-1} \times$ (5, 6, 9, 12, 15, 18, 21, 24, 27, 30) for the 4.86~GHz map,
0.20~mJy~beam$^{-1} \times$ (3, 4, 5, 7, 9, 11, 13, 15, 20) for the 8.44~GHz map, 
and 0.013~mJy~beam$^{-1} \times$ (3, 6, 9, 15, 21, 27, 33, 39, 45, 51) for the 133~GHz map. 
}\label{fig:radio}
\end{center}
\end{figure*}

\subsection{Radio Continuum Emission: Existence of Radio Jet-Lobe}\label{sec:radio}

The Band~4 continuum observation of Arp~187 was carried out 
using the ALMA 12-m array on 2016 July 25 (Cycle~3; ID = 2015.1.01005.S).  
We used four SPectral Windows (SPWs) with Time Domain Mode.  
The center frequency of the four SPWs is 133~GHz, 
and the total bandwidth is 8~GHz.  
The number of 12-m antennas was 36.  
The baseline lengths range from 15~m to 1124~m.  
Data calibration and imaging were carried out 
using the Common Astronomy Software Applications package (CASA, ver. 4.5.3).  
We used the delivered calibrated $uv$ data 
and made the continuum map by clipping the visibility ($uv$ distance $\geq 10$k$\lambda$).
The synthesized beam size is $0\farcs46 \times 0\farcs39$ 
(position angle ${\rm (PA)} = -82\fdg7$) 
by adopting Briggs weighting of the visibility $({\rm robust} = 0.5)$, 
and the rms noise level is 0.013~mJy~beam$^{-1}$.
We regard the accuracy of the absolute flux calibration 
as 5\% throughout this paper according to the ALMA Cycle~3 Technical 
Handbook\footnote{A. Remijan et al., 2015, ALMA Cycle 3 Technical Handbook Version 1.0, ALMA}.

In addition, we used archival calibrated $uv$ data 
obtained with VLA C- and X-bands,
and made the continuum maps using CASA.
We clipped the visibilities before imaging 
so that all the VLA and ALMA data have the same shortest UV range.
The synthesized beam size of the 4.86~GHz map 
is $0\farcs43 \times 0\farcs34$ $({\rm PA} = -4\fdg9)$
by adopting uniform weighting of the visibility,
and the rms noise level is 0.28~mJy beam$^{-1}$.
The synthesized beam size of the 8.44~GHz map 
is $0\farcs36 \times 0\farcs22$ $({\rm PA} = 34\fdg1)$ 
by adopting Briggs weighting of the visibility $({\rm robust} = 0.5)$
and the rms noise level is 0.20~mJy~beam$^{-1}$.

The radio continuum maps are shown in Figure~\ref{fig:radio}.  
They clearly show the structure of the jet lobes, 
which are located at both sides of the nucleus.  
The projected distance between the lobe cores is $\sim 5$~kpc.  
We estimate the kinematic age of the lobes, 
assuming the jet angle to the line of sight of $90\arcdeg$ 
and a typical expansion speed of radio lobes 
\citep[$0.1c$; e.g.,][]{mur99,nag06}. 
This gives the kinematic age of $\sim8 \times 10^4$~years.

We also confirm no additional radio lobes in the field of view (FoV) 
 of the VLA and ALMA observations.
The FoV of the VLA observation at 4.86~GHz is 9.3~arcmin,
which is the largest among the three observations.
This corresponds to the physical size of 221~kpc in radius.
If there are radio lobes outside of the FoV, the
lower limit of their kinematic age is $6.6 \times 10^6$~years. 
Thus Arp~187 has not launched the jet over $6.6 \times 10^6$~years
before the current 5~kpc size one, or the larger radio lobes, if any, 
already become undetectable due to rapid energy loss \citep[e.g.,][]{god17}.


\begin{figure}
\begin{center}
\includegraphics[width=8.0cm]{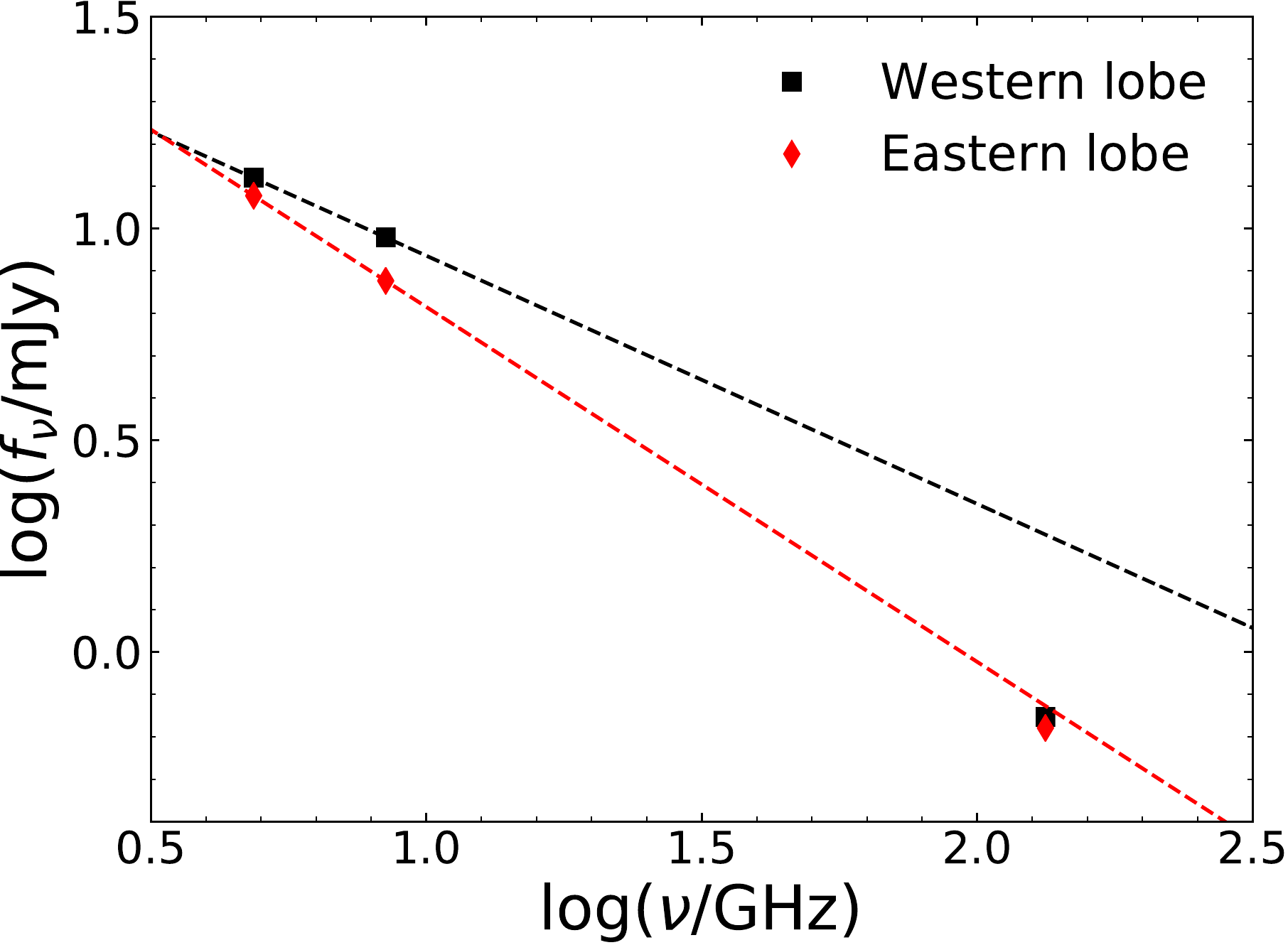}~
\caption{
The SEDs of the jet lobes.
The spectral indexes estimated from two points at 4.86~GHz and 8.44~GHz are 
$\log f_{\nu} = 9.2 - 0.84 \log \nu$ for the western lobe (black) 
and $\log f_{\nu} = 6.8 - 0.59 \log \nu$ for the eastern lobe (red).
The errors of the flux densities are smaller than the size of symbols.
}\label{fig:radioSED}
\end{center}
\end{figure}

\begin{deluxetable}{ccccc}
\tablecaption{Flux density of the jet lobes\label{tab:radio}}
\tabletypesize{\footnotesize}
\tablewidth{0pt}
\tablehead{
&\multicolumn{2}{c}{Eastern lobe}&\multicolumn{2}{c}{Western lobe}\\
\colhead{Frequency}&
\colhead{Peak}&
\colhead{Total}&
\colhead{Peak}&
\colhead{Total}
}
\startdata
4.86~GHz&12.0&190&13.2&160\\
8.44~GHz&7.53&92&9.54&100\\
133~GHz&0.662&5.3&0.704&7.2\\
\enddata
\tablecomments{
The unit of the flux density is mJy.
These values are measured in the beam-matched images ($\theta$ = 0\farcs47).}
\end{deluxetable}

\begin{figure*}
\begin{center}
\includegraphics[width=0.42\linewidth]{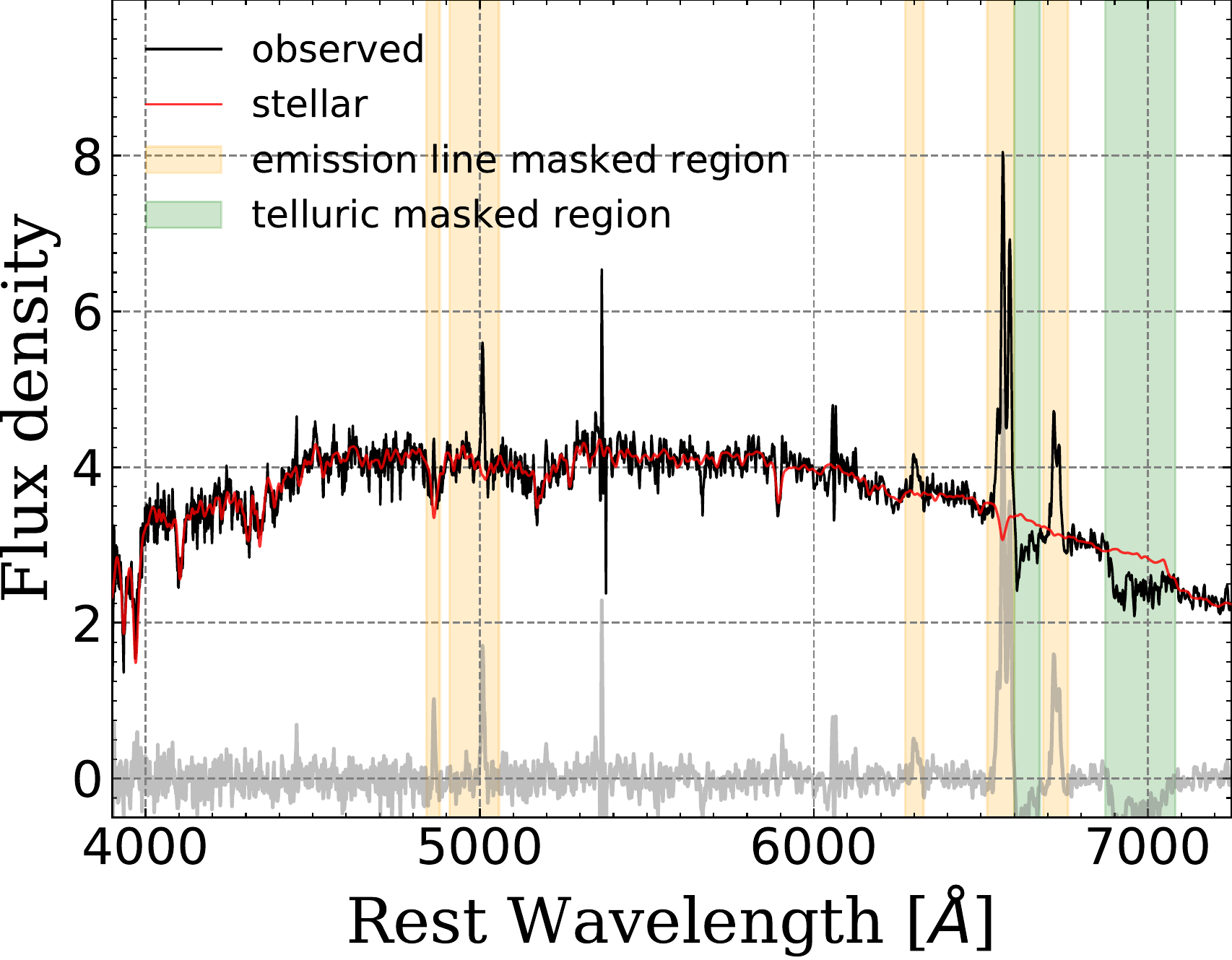}~
\includegraphics[width=0.47\linewidth]{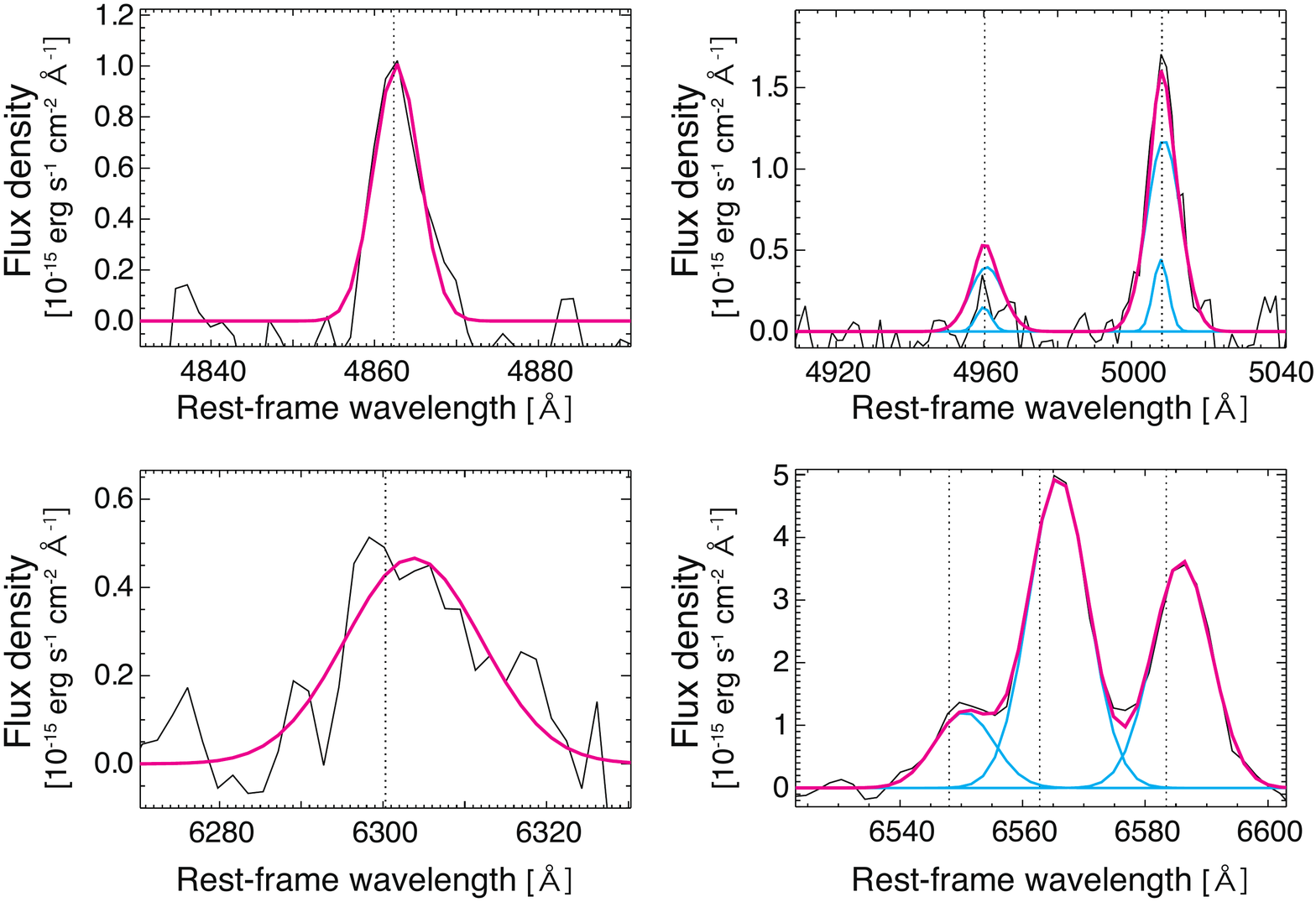}
\caption{
(Left): The optical spectrum of Arp~187 (solid black line) obtained from 6dF galaxy survey.
X-axis shows the rest-frame wavelength with the unit of \AA.
Y-axis shows the flux density with the unit of $10^{-15}$~erg~s$^{-1}$~cm$^{-2}$~{\AA}$^{-1}$.
The best-fit stellar spectrum is also shown in the red solid line.
The starlight subtracted spectrum is shown in gray line at the zero flux level.
The emission line and the telluric masked region is shown in the orange and green, respectively.
(Right): The spectra and the best-fit models for the Balmer lines, the
[\ion{O}{3}]-doublet ($\lambda4959,\lambda5007$), 
the [\ion{O}{1}]$\lambda6300$, 
and the [\ion{N}{2}]-doublet lines.
(upper-left: H$\beta$, upper-right: [\ion{O}{3}]$\lambda4959,\lambda5007$
doublet, bottom-left:[\ion{O}{1}]$\lambda6300$, bottom-right: H$\alpha$ and [\ion{N}{2}]-doublet).
The total model spectrum is shown as pink solid line.
The single Gaussian line component is shown with cyan solid line (for
H$\beta$, H$\alpha$, and the [\ion{N}{2}]-doublet),
while the double Gaussian line component is shown with two cyan lines
(for the [\ion{O}{3}] doublet) as discussed in Section~\ref{sec:6dFspec}.
}\label{fig:6dFspec}
\end{center}
\end{figure*}

We also convolved these maps to $0\farcs47$ angular resolution
to make Spectral Energy Distributions (SEDs) of the jet lobes.
The peak and total flux densities of the eastern and western robes 
are summarized in Table~\ref{tab:radio}.
Each peak flux density is measured at the emission peak at 4.86~GHz,
using the beam-matched maps.  
The radio SEDs of the lobes are shown in Figure~\ref{fig:radioSED}.
The spectral index $\alpha$ was estimated from two points at 4.86~GHz and 8.44~GHz 
by using a fitting function of $f_\nu\propto\nu^\alpha$.
The SED shows a steep spectral index of $\alpha\approx-0.59$ for 
the eastern lobe (the red dashed line), 
and the flux point at 133~GHz is nearly on the extrapolated line
from the VLA bands.
This is consistent with optically thin synchrotron radiation 
that is expected for the jet emission.
On the other hand, the flux density of the western lobes
at 113~GHz is smaller than what is expected 
from the flux densities at 4.86~GHz and 8.44~GHz, 
assuming that the SED can be fitted by $f_\nu\propto\nu^\alpha$
(the black dashed line).
This suggests the spectral aging for the western jet 
\citep[e.g.,][]{jam08,sai09},
while the exact age cannot be determined based on the current SED sampling.

There is no clear emission in the nucleus (jet core) at 4.86~GHz and 8.44~GHz, 
although the faint emission is seen at 133~GHz.
This is not due to the dynamical range limit of VLA.  
The 3$\sigma$ upper limits of the flux densities 
measured in the non beam-matched images 
are 0.84~mJy at 4.86~GHz and 0.60~mJy at 8.44~GHz.  
Assuming the spectral index of the jet core of $\alpha = -0.5$,
the 133~GHz flux density is expected to be 0.16~mJy 
based on the upper limit of the 8.44~GHz flux density.  
However, the observed 133~GHz flux density is $\leq0.052$~mJy, 
which is three times lower than that extrapolated from the 8.44~GHz flux density.  
The non-detection of the continuum emission at 4.86~GHz and 8.44~GHz 
cannot be explained by free-free absorption 
which causes the flux excess at high-frequency.  
Thus, the non-detection at several GHz and 
strong constraint on the 133~GHz flux density
lead to the conclusion of the presence of a significantly faint core in Arp~187.

\subsection{6dF Optical Spectra}\label{sec:6dFspec}

We perform the spectral measurements to investigate the properties of the NLR.
We first obtained the archival optical spectrum of Arp~187 from the 6dF galaxy survey \citep{jon09}. The spectrum covers a range from 3900\AA~to 7500\AA~with a 
fiber aperture of 6.7~arcsec (equivalent to $5.3$~kpc).
Since the 6dF spectra are not flux calibrated on a nightly basis,
we have normalized the spectrum based on the 6dF optical photometry at the $R$ band
of $R=14.33$~mag.

Figure~\ref{fig:6dFspec} shows the optical spectrum of Arp~187
in the left panel (solid black line).
The observed spectrum does not show any features of
the big blue bump originating from the AGN accretion disk, 
nor the broad emission lines \citep[e.g.,][]{mal82,van01}. 
Thus, we conclude that Arp~187 is at least not a type-1 AGN.
 
We then perform the spectral fitting to obtain the properties of the NLR.
For the spectral fitting, we follow the routine in 
\cite{bae17} and later \cite{tob17a}, 
where they have performed the spectral fitting to the SDSS spectra.
We first subtract the stellar continuum from the spectrum
using a best-fit stellar template based on the wide range of the stellar
population models \citep[MILES;][]{san06} with solar metallicity
and the age spanning from 60~Mys to 12.6~Gyr.
In order to obtain the qualified stellar continuum, 
we mask the strong emission lines of H$\beta$, [\ion{O}{1}], [\ion{N}{2}],
H$\alpha$, and [\ion{S}{2}] with the masking width of 1300~km~s$^{-1}$,
which is corresponding to ${\rm FWHM}\sim 1000$~km~s$^{-1}$.
We also mask the [\ion{O}{3}] doublet with the masking width
of 3000~km~s$^{-1}$ in order to avoid the possible contamination
from the strong outflow originated from the [\ion{O}{3}] emission lines.
In addition, since some of the telluric absorptions are not well removed, 
we also masked those wavelength bands from the stellar spectral fitting.

From the starlight-subtracted spectrum as shown in gray line
at the zero flux level in Figure~\ref{fig:6dFspec}, we then fit the H$\alpha$, H$\beta$,
[\ion{N}{2}]-doublet, [\ion{O}{1}]$\lambda6300$
using a single
Gaussian function and [\ion{O}{3}] doublet ([\ion{O}{3}]$\lambda4959, 5007$) 
with double-Gaussian function using the IDL/MPFIT code,
which is a $\chi^2$-minimization routine \citep{mar09}. 
We assume that the [\ion{O}{3}] doublet and the other narrow-lines
have independent kinematics, while the [\ion{O}{3}] doublet
has the same velocity and velocity dispersion as each other.
The observed spectrum (black line) as well as the fitting results
 is compiled for the entire spectral range (Figure~\ref{fig:6dFspec}, left panel)
 and for each lines (Figure~\ref{fig:6dFspec}, right panel).

\subsubsection{Existence of NLR}
\begin{figure}
\begin{center}
\includegraphics[width=8.3cm]{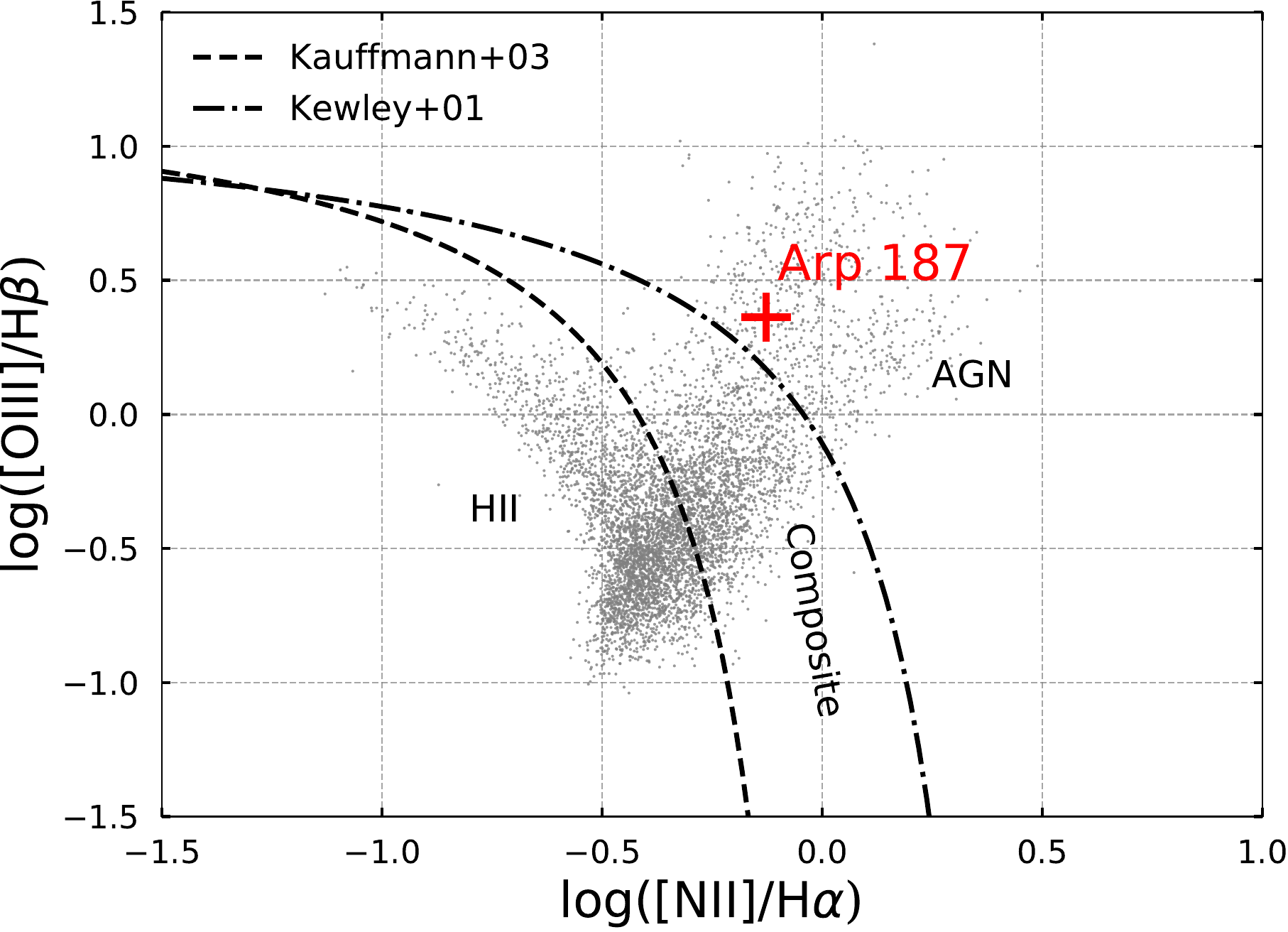}~
\caption{
Classification of the galaxies into AGN, composite galaxies, and
\ion{H}{2} regions
 using line diagnostics diagrams \citep{kew01,kew06}.
The dot-dashed/dashed line represents the
relation of \cite{kew01}/\cite{kau03}, respectively. 
Arp~187 is located in the locus of AGN (red cross).
Gray dots represent the data points of SDSS DR7 galaxies \citep{aba09}.
}\label{fig:BPT}
\end{center}
\end{figure}

 We first apply the emission-line diagnostics, which give a separation between
 the NLR ionized by AGN and the \ion{H}{2} region in the starburst galaxies
 \citep[][]{vei87,kew06}. 
Figure~\ref{fig:BPT} shows that Arp~187 is 
 classified as AGN. Thus, it shows that Arp~187 hosts the NLR, 
 one of the large scale AGN indicators.
 
Since the diagnostic shows that [\ion{O}{3}]$\lambda5007$ (hereafter, [\ion{O}{3}]) 
emission is dominated from the NLR,
we then measure the [\ion{O}{3}] luminosity ($L_{\rm [OIII]}$), 
and [\ion{O}{1}]$\lambda6300$ (here after, [\ion{O}{1}]) luminosity ($L_{\rm [OI]}$) 
to estimate the AGN bolometric luminosity because
$L_{\rm [OIII]}$ alone \citep[e.g.,][]{hec04,hec05,lam13,ued15}
or the combination of $L_{\rm [OIII]}$ and 
$L_{\rm [OI]}$ \citep[e.g.,][]{net09} is often used as a proxy for AGN power.
The observed [\ion{O}{3}] and [\ion{O}{1}] luminosities are
$L_{\rm [OIII]}=6.2 \times 10^{40}$~erg~s$^{-1}$
and $L_{\rm [OI]}=3.8 \times 10^{40}$~erg~s$^{-1}$.
We also calculate an extinction-corrected, 
intrinsic luminosity of [\ion{O}{3}] ($L_{\rm [OIII]}^{\rm int}$)
and [\ion{O}{1}] ($L_{\rm [OI]}^{\rm int}$)
luminosities  from the Balmer decrement of H$\alpha$/H$\beta$
\citep[e.g,.][]{cal94,dom13}.
The values are
$L_{\rm [OIII]}^{\rm int}=3.2 \times 10^{42}$~erg~s$^{-1}$ and
$L_{\rm [OI]}^{\rm int}=2.0 \times 10^{42}$~erg~s$^{-1}$.
We calculate the AGN bolometric luminosity $\lbol$ 
using the both lines \citep{net09,mat15b} by
\begin{equation}
\log L_{\rm bol} = 3.8 + 0.25 \log L_{\rm [OIII]}^{\rm int} + 0.75 L_{\rm [OI]}^{\rm int}.
\end{equation}
The estimated value is $\log (\lbol/{\rm erg}~{\rm s}^{-1}) = 46.2$,
which reaches to the typical luminosity of the SDSS quasars at $z\sim1$--2
\citep[e.g.,][]{she11}.

\subsubsection{NLR Size}
The size of the NLR is also an important indicator 
to estimate the fading timescale of AGN for Arp~187.
Since the integral field unit (IFU) observations are still not available, we estimate the NLR size from the empirical
relationship between the [\ion{O}{3}] emission size and the [\ion{O}{3}] luminosity.

The [\ion{O}{3}] emission sizes have been measured from either narrowband imaging \citep{ben02,sch03},
 long-slit spectroscopy \citep{fra03,ben06,gre11,hai13}, or the IFU observations \citep{hum10,hus13,liu13,kar16,bae17}.
The narrowband imaging is more often used for studying the NLR morphology, while
the long-slit spectroscopy and IFU has an advantage to reach shallower sensitivity limits.

We first estimate the NLR sizes ($R_{\rm NLR}$)
using the size-luminosity relations of \cite{bae17}
obtained from the IFU observations of nearby type-2 Seyferts and quasars.
This relation has two advantages; 1)
their study uses the extinction-uncorrected [\ion{O}{3}] luminosity
for the size-luminosity relation to reduce the uncertainty of the
extinction correction to estimate the intrinsic [\ion{O}{3}] luminosity,
and 2) they estimate the NLR size based on the line diagnostic diagrams
discussed in Section~\ref{sec:6dFspec}, which pick up the region
where the [\ion{O}{3}] emission is dominated from AGN.
The size-luminosity relation of \cite{bae17} is given by

\begin{equation}
\log \left(\frac{R_{\rm NLR}}{\rm pc} \right)=
0.41 \times \log \left( \frac{L_{\rm [OIII]}}{{\rm erg}~{\rm s}^{-1}}\right)  - 14.00.
\end{equation}
The estimated NLR size is $R_{\rm NLR}\simeq 530$~pc.

We also estimate the [\ion{O}{3}] emitting sizes ($R_{\rm [OIII]}$) using 
the size-luminosity relations from the literature.
Note that the relation is based on the [\ion{O}{3}] emission size, 
therefore the estimated size should be the upper bound of 
the NLR size since the NLR should fulfill the certain emission line ratios
as shown in Figure~\ref{fig:6dFspec}.
We apply the relation based on the 
narrow-band imaging observations of quasar population \citep{sch03} by
\begin{equation}
\log \left(\frac{R_{\rm [OIII]}}{\rm pc} \right) = 0.42 \times \log \left( \frac{L_{\rm [OIII]}^{\rm int}}{{\rm erg}~{\rm s}^{-1}}\right) -14.72.
\end{equation}
This gives $R_{\rm NLR} < R_{\rm [OIII]}\simeq 1.4$~kpc,
which is $\sim2.5$ times larger than $R_{\rm NLR}$ obtained above.
This result is in good agreement with the 
previous IFU studies \citep{kar16,bae17}, who reported that
$R_{\rm [OIII]}$ is on average a few times larger than $R_{\rm NLR}$.
In this study, we assume $R_{\rm NLR}=530$--$1400$~pc
as the possible NLR size range.

\subsection{\textit{Spitzer}/IRS Spectra}\label{sec:Spitzer}

Since the detailed spectral analysis of Arp 187 obtained from the \textit{Spitzer}/InfraRed Spectrograph (IRS) is compiled in \cite{ich16},
here we provide a brief summary of the findings.
We have found that the thermal emission from the AGN is already weak in Arp~187,
with an upper bound of the 12~$\mu$m luminosity of $L_{\rm 12{\mu}m} < 1.5\times10^{42}$~erg~s$^{-1}$.
This is equivalent to $L_{\rm bol}< 2.2 \times 10^{43}$~erg~s$^{-1}$ using the relation of $L_{\rm 12{\mu}m}$ and 14--195~keV luminosities \citep{ich17a} and the bolometric correction of $L_{\rm bol}/L_{14-195}=8.47$ \citep{ric17a}.
The spectrum of Arp~187 shows a marginal detection of the
[\ion{O}{4}]~25.89~$\mu$m line at S/N$\sim3$,
which is also widely used AGN indicator.
The [\ion{O}{4}] luminosity is obtained with $L_{\rm [OIV]}=6.7 \times 10^{40}$~erg~s$^{-1}$,
which is equivalent to $L_{2-10}=3.0 \times 10^{43}$~erg~s$^{-1}$ based on the luminosity relations obtained by \cite{lam10}.
Since the ionization potential of [\ion{O}{4}] is higher 
($E_{\rm p}=54.9$~eV) than other typical 
NLR indicators including [\ion{O}{3}] line \citep[e.g.,][]{rig09},
the emission size is expected to be smaller than that of [\ion{O}{3}]
\citep{kom08,tob17a}, which is 0.5--1.4~kpc in this study.
Thus, we take the emission size of $<1.4$~kpc as a fiducial value.

\begin{figure}
\begin{center}
\includegraphics[width=8.0cm]{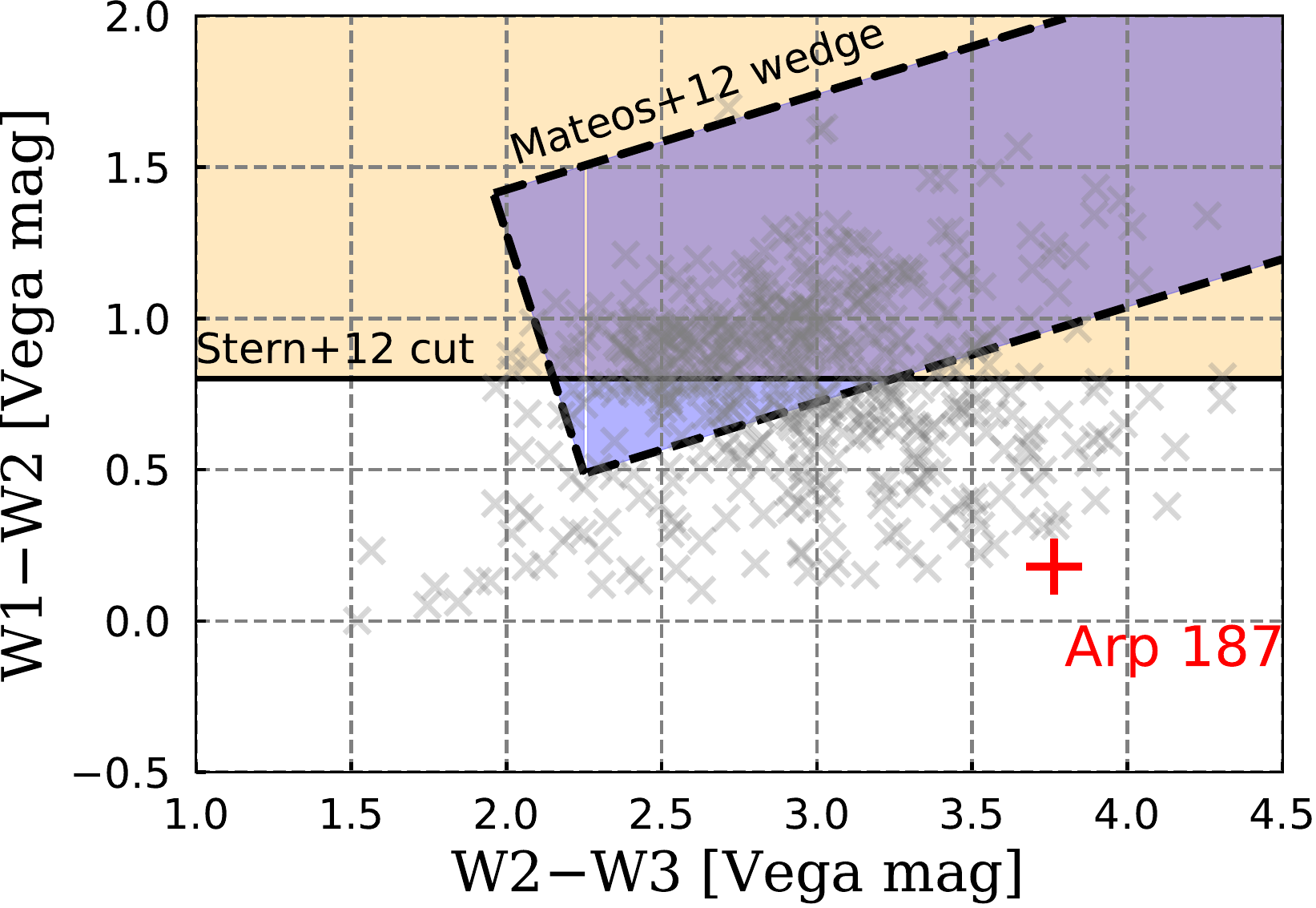}~
\caption{
\textit{WISE} W1 (3.4~$\mu$m)--W2 (4.6~$\mu$m) versus 
W2 (4.6~$\mu$m)--W3 (12~$\mu$m) two color diagram in the
unit of Vega magnitude.
The red and gray cross represents Arp 187 and the \textit{Swift}/BAT AGN sample in the local universe, respectively.
The orange and purple area represents the AGN region proposed
by \cite{ste12} and \cite{mat12}, respectively.
}\label{fig:WISEcolor}
\end{center}
\end{figure}

\begin{figure*}
\begin{center}
\includegraphics[width=0.47\linewidth]{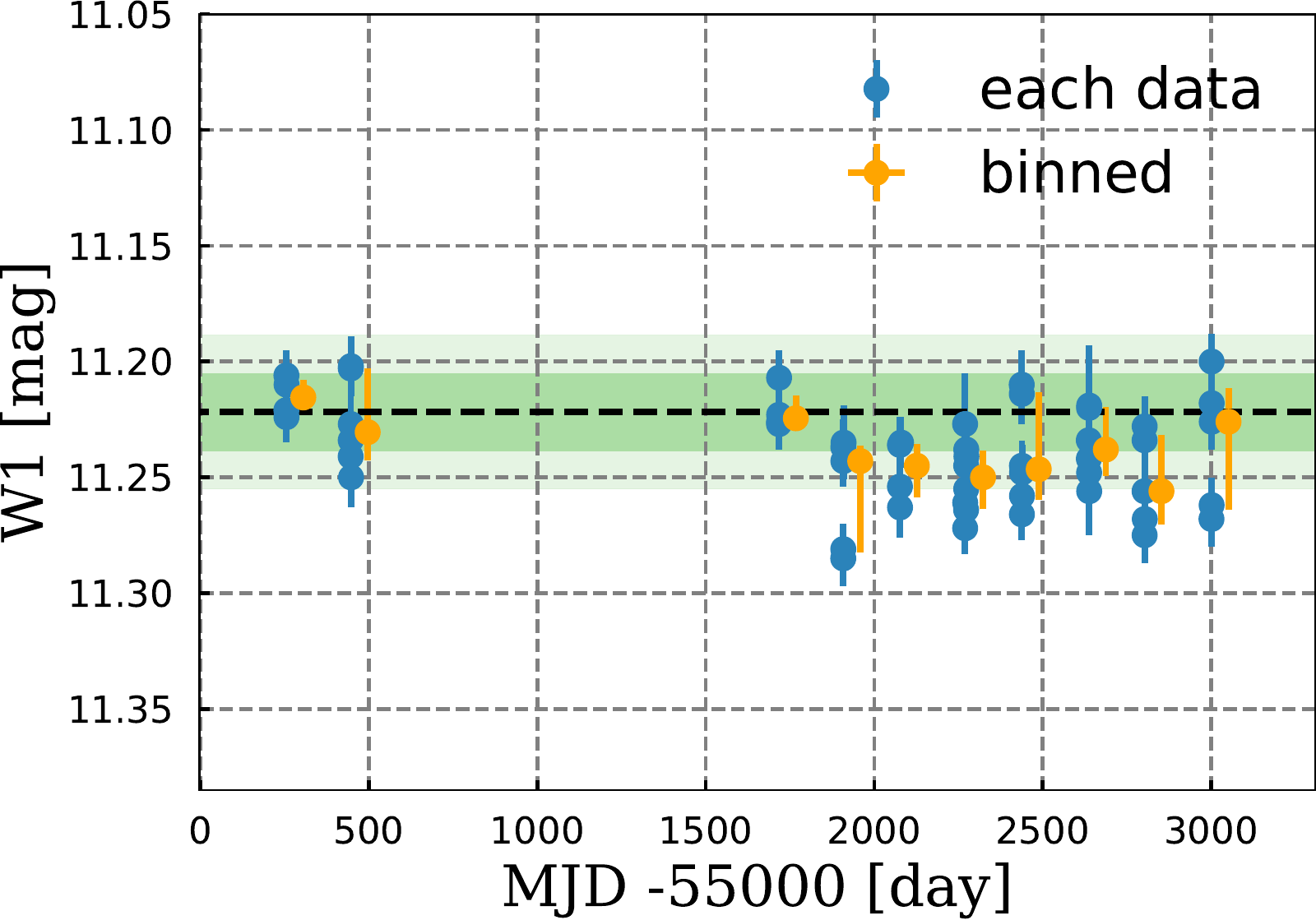}~
\includegraphics[width=0.47\linewidth]{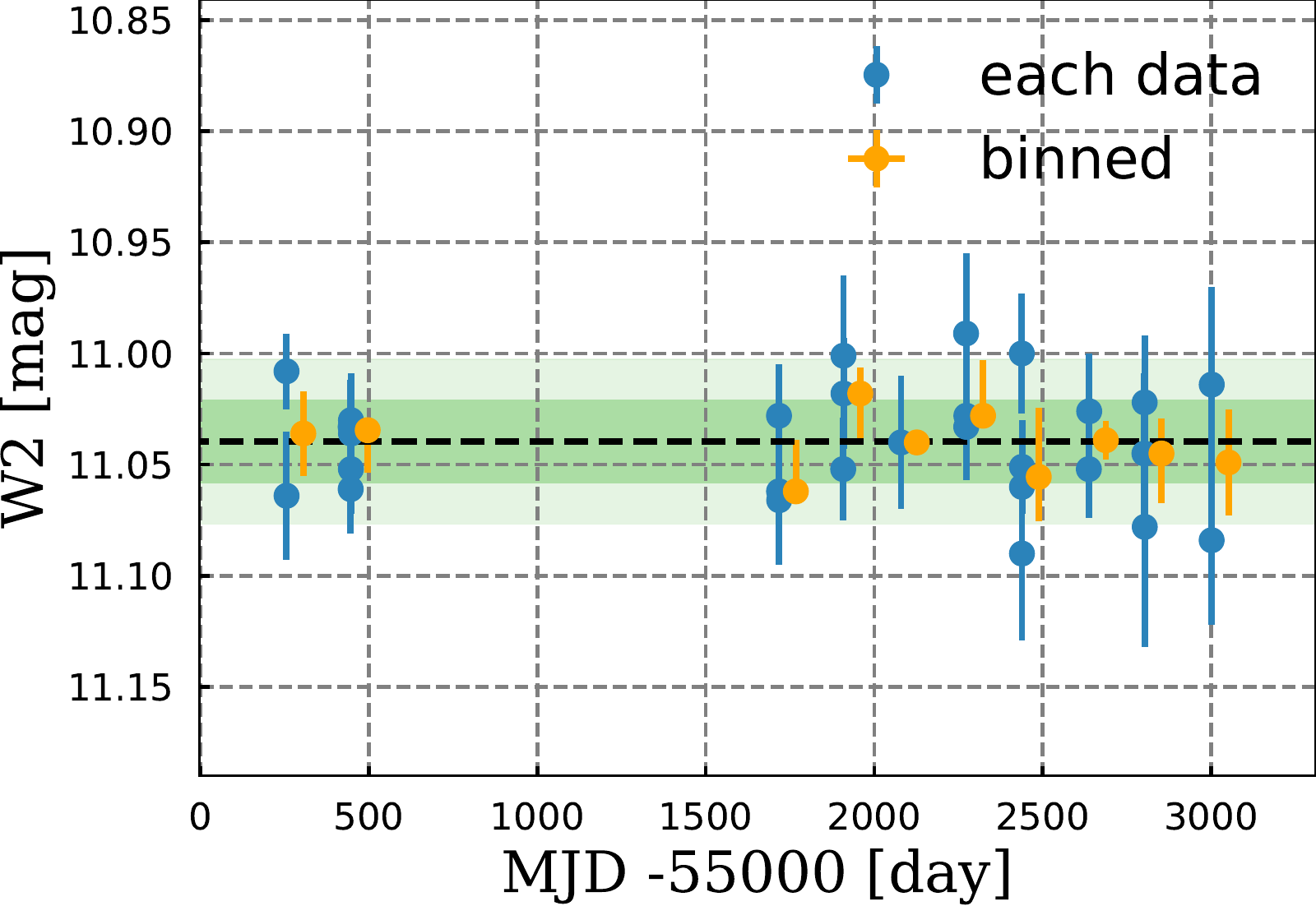}
\caption{
Infrared light curves of Arp~187 in the W1 and W2 band.
The single exposures with the error are shown in blue, 
while the median value in each epoch is
shown in orange circle with the error bars showing 
the inter-percentage range with 68\% of the sample.
The orange circles are shifted to 50 days after the real values for clarity.
The dashed line with green shade is the flux obtained from
\textsc{Allwise} with the 1$\sigma$ (dark green) and 2$\sigma$ (light green) scatter, 
representing the average magnitude in the \textsc{Allwise} epoch.
}\label{fig:WISElightcurve}
\end{center}
\end{figure*}

\subsection{WISE Colors}\label{sec:WISEcolor}
IR color-color selection is useful to identify AGN candidates 
using the feature of the MIR bump originated from the AGN torus.
Figure~\ref{fig:WISEcolor} shows the location of Arp~187 on the \textit{WISE} color-color plane.
It is known that increasing levels of AGN contribution to the MIR emission moves sources
upwards in the color-color plane with the color cut of $W1-W2>0.8$ \citep[orange area;][]{ste12}
and also within the AGN wedge \citep[blue area;][]{mat12}.
The figure clearly shows that Arp~187 does not fulfill either of the criteria above,
indicating either 1) Arp~187 does not host AGN or 2) the AGN activity is quite weak even if it exists
\citep[on average $L_{\rm 14-195}<10^{43}$~erg~s$^{-1}$, as suggested by][]{ich17a}.

\subsection{WISE IR Time Variability}\label{sec:WISElightcurve}

It is worth to trace the IR luminosity drop from the AGN torus after the AGN quenching.
Since Arp~187 is bright enough to be detected in the single exposure with the \textit{WISE} mission,
we have collected both cryogenic (\textit{WISE} All-sky database) and 
post-cryogenic multi-epoch photometry (\textit{WISE}
3~band and Post-Cryo database) 
from the \textsc{Allwise} \citep{wri10,mai11} covering the observation
between 2010 January and 2011 February (${\rm MJD}-55000=200$--$600$),
and the most recent \textsc{Neowise} \citep{mai14} data release 2018 covering the observation between 2013 December 13 and 2017 December 13, UTC 
(${\rm MJD}-55000=1600$--$3100$).
\textit{WISE} has 90-min orbit and conducts $\approx 12$ observations
of a source over a $\approx 1$~day period, and a given location is
observed every six months.

In this study we used standard aperture magnitude (\texttt{w1/2mag}).
We applied a cross-matching radius of 2 arcsec, 
based on the positional accuracy with the 2MASS catalog \citep[see also][]{ich12, ich17a}.
After this matching, 26 and 122 data-points were obtained from the \textsc{Allwise} and \textsc{Neowise} epoch, respectively.
Then we select good quality single-epoch data points based on
the good quality frame score (\texttt{qual\_frame>0} and \texttt{qi\_fact>0}),
locating the enough distance from the South Atlantic Anomalies (\texttt{saa\_sep>0}),
and avoiding the possible contamination from the moon (\texttt{moon\_masked=0}).
This reduces the sample into 24 (\textsc{Allwise}) and 
109 (\textsc{Neowise}), respectively.
Finally, we applied the aperture measurement quality flag (\texttt{w1/2flg=0})
in order to avoid the contamination in the aperture.
The final data points are  10 (W1) and 8 (W2) for \textsc{Allwise}
43 (W1) and 21 (W2) for \textsc{Neowise}.
All the data points fulfill the flux quality \texttt{ph\_qual=A},
with a signal-to-noise ratio larger than 10.0.
We also checked sources of contamination and/or biased flux, 
due to proximity to an image artifact (e.g., diffraction spikes, scattered-light halos,
and/or optical ghosts), using the contamination flag \texttt{cc\_flags}.
All the data points are \texttt{cc\_flags=0}, that are unaffected by known artifacts.

Figure~\ref{fig:WISElightcurve} shows the light curve of W1 (3.4~$\mu$m) and W2 (4.6~$\mu$m).  
The light curves in the W1 and W2 span a baseline of roughly 2800 days $\simeq 7.0$~years. Each observation is shown in blue and a binned observations
within one-day is also shown in orange.
As shown in Figure~\ref{fig:WISElightcurve}, 
no clear variability is detected in the \textsc{Allwise} and \textsc{Neowise} epoch,
and also between the two epochs.
Actually, the \textsc{Allwise} catalog provides a variability flag (\verb|var_flag|) and
its value is $\verb|var_flag|=0$, suggesting that the significant variability between different
exposures are not detected during the \textsc{Allwise} survey, which shows the consistent result.

\begin{figure}
\begin{center}
\includegraphics[width=8.0cm]{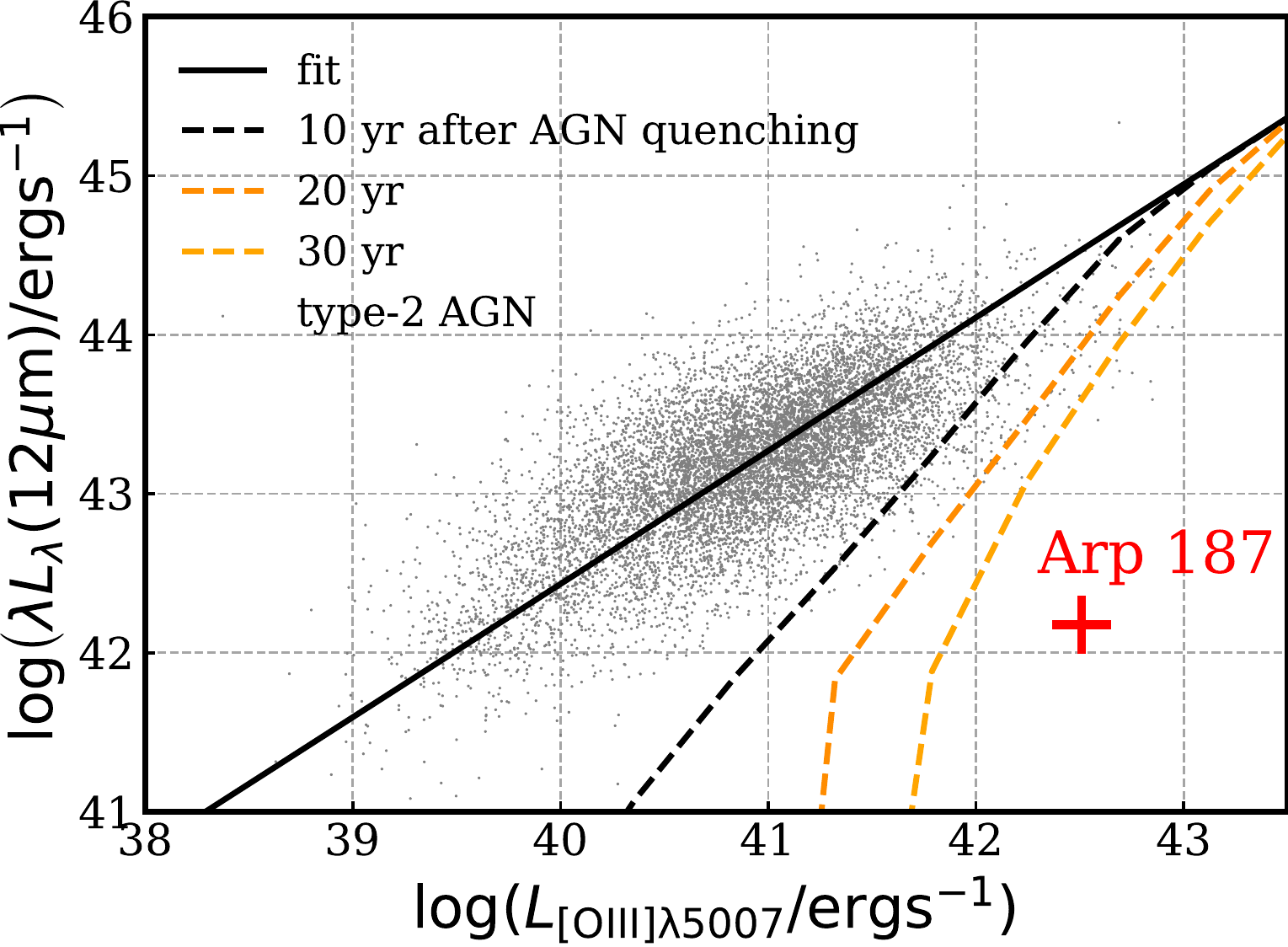}~
\caption{
Correlation between the [\ion{O}{3}]$\lambda5007$ and 12~$\mu$m luminosities. 
The black solid line represents the regression line obtained from the bisector fitting of the
type-2 AGN sample from SDSS DR12 galaxies (gray dots).
The dashed lines represent the time evolution of the AGN luminosity relation
the [\ion{O}{3}]$\lambda5007$ and 12~$\mu$m after AGN quenching \citep{ich17c}; 10~years (black),
20~years (dark-orange), and 30~years (orange).
The location of Arp~187 is shown in red cross.
}\label{fig:L12vsO3}
\end{center}
\end{figure}

\subsection{Relationship between 12~$\mu$m and [\ion{O}{3}] Luminosities}\label{sec:L12vsO3}

It is widely known that both 12~$\mu$m and the
[\ion{O}{3}] luminosities are good proxies for AGN power,
and it is a natural consequence that there is a luminosity 
correlations between 12~$\mu$m and the [\ion{O}{3}]
luminosities \citep[e.g.,][]{tob14}. 

Figure~\ref{fig:L12vsO3} shows the
 relationship between 12~$\mu$m and [\ion{O}{3}]$\lambda5007$
luminosities using the SDSS DR12 type-2 AGN sample with the cross-matching by
 the \textit{WISE}~W3 (12~$\mu$m) band. As expected, two AGN indicators have
 a nice luminosity correlation. However,
Arp~187 does not follow the luminosity relation and is located at the
  right bottom in the plane, suggesting that Arp~187 is in the locus of dying AGN.
 \cite{ich17c} estimated the typical cooling timescale of the dusty torus once the AGN is suddenly quenched.
 The thermal MIR dust emission from AGN should decay times of 10--100~years, mainly depending on the
 time-lag caused by the light travel time from the nucleus to the dust sublimation radius,
 while the $L_{\rm [OIII]}$ remains the same luminosity
 for over $>10^3$~years because of its larger physical size.
\cite{ich17c} also calculated how those dying AGN move in the luminosity--luminosity plane of the two AGN indicators,
and showed that those dying AGN should locate at the bottom right of the luminosity relation.
The location of Arp~187 in Figure~\ref{fig:L12vsO3} is consistent with the idea, and
below the relationship of the line after 30~years of AGN quenching 
(the orange dashed line).

\begin{deluxetable*}{lcCccccc}
\tablecaption{AGN indicators and their estimated luminosities \label{tab:AGNindicators}}
\tabletypesize{\footnotesize}
\tablecolumns{8}
\tablenum{1}
\tablewidth{0pt}
\tablehead{
\multicolumn{8}{c}{Large scale AGN indicators ($>100$~pc)}\\
\hline
\colhead{Type of AGN indicators} &
\colhead{AGN sign} &
\colhead{$\log (L_{\rm AGN}/{\rm erg}~{\rm s}^{-1})$} &
\colhead{$\log (L_{\rm bol}/{\rm erg}~{\rm s}^{-1})$} &
\colhead{$\log \lambda_{\rm Edd}$} & 
\colhead{Size (radius)} & 
\colhead{$t_{\rm retro}\tablenotemark{a}$}  &
\colhead{Reference}
}
\startdata
jet lobe & $\checkmark$ & $\cdots$  &  $\cdots$ & $\cdots$
& $2.5$~kpc & $8.1 \times 10^{4}$~yr & Section~\ref{sec:radio}\\
NLR ([OIII]$\lambda5007$) & $\checkmark$ & $\log L_{\rm [OIII]}^{\rm int}=42.51$  &
 $46.15$ & $-0.77$ &  $0.5$--$1.4$~kpc & $(1.7$--$4.6) \times 10^{3}$~yr & Section~\ref{sec:6dFspec}\\
NLR ([OIV]25.89~$\mu$m) & $\checkmark$ & $\log L_{\rm [OIV]}^{\rm int}=40.83$ &
$44.78$  & $-2.15$ & $\leq 1.4$~kpc & $\leq 4.6 \times 10^{3}$~yr & Section~\ref{sec:Spitzer} \\
\hline
\multicolumn{8}{c}{Small scale AGN indicators ($<10$~pc)}\\
\hline
Dust torus (\textit{Spitzer}/IRS Spec) & \text{\sffamily X} & $\log L_{\rm 12 {\mu}m}<42.18$ &
 $<43.34^{a}$ & $<-3.59$ & $\simeq 10$~pc & $10$--$100$~yr & Section~\ref{sec:Spitzer} \\
Dust torus (\textit{WISE} IR colors) & \text{\sffamily X} & $\cdots$ & $\cdots$ & $\cdots$ &
$\cdots$ & $10$--$100$~yr & Section~\ref{sec:WISEcolor} \\
Dust torus (\textit{WISE} IR light curve) & \text{\sffamily X} & $\cdots$ & $\cdots$ & $\cdots$ &
$\cdots$ & $10$--$100$~yr & Section~\ref{sec:WISElightcurve} \\
Dust torus ($L(\rm [OIII])$ vs. $L_{\rm 12 {\mu} m}$) &  \text{\sffamily X} & $\log L_{\rm 12 {\mu}m}<42.18$ &
 $<43.34^{a}$ &  $<-3.59$ & $\simeq10$~pc & $>30$~yr & Section~\ref{sec:L12vsO3}\\
jet core & \text{\sffamily X} & $\log L_{5 {\rm GHz}}<37.57$ & $\cdots$ & $\cdots$ & $\leq 1$~pc 
& Current & Section~\ref{sec:radio} \\
BLR (optical Spec) & \text{\sffamily X} & $\cdots$ & $\cdots$ & $\cdots$ &  $<0.1$~pc & Current (1--10~yr) & Section~\ref{sec:6dFspec}\\
Electron corona (X-ray) &  \text{\sffamily X} & $\log L_{2-10}<43.44$ & $<44.38^{a}$ & $<-2.55$ & $\ll 0.1$~pc & Current & Section~\ref{sec:X-ray} \\
\hline
\multicolumn{8}{c}{Other Relations}\\
\hline
BH fundamental plane (FP) & $\cdots$ & $\log L_{2-10}< 39.60$ & $<40.90^{a}$ & $<-6.03$ & $\cdots$ &  Current &
Section~\ref{sec:BHfundamental} \\
\enddata
\tablecomments{The list of AGN indicators
and the retrospective time ($t_{\rm retro}$) of the AGN indicator.
Except the jet lobe, the all time is the
light crossing time with the definition of $t_{\rm lc}={\rm size}/c$.
\tablenotetext{a}{From the X-ray luminosities, we use the bolometric correction of $L_{\rm bol}/L_{2-10}=20$ \citep{ric17a}.
Then, under the assumption of the photon index of $\Gamma=1.8$ \citep{ric17}, 
we use the conversion among the X-ray luminosities
of $L_{2-10}/L_{14-195}=0.42$ and $L_{2-10}/L_{0.5-2}=1.57$.
We also use the luminosity relations between $L_{\rm 12{\mu}m}$ and $L_{14-195}$ of 
$\log (L_{\rm 12{\mu}m}/10^{43}~{\rm erg}~{\rm s}^{-1})=-0.21+1.05 \log (L_{14-195}/10^{43}~{\rm erg}~{\rm s}^{-1})$
to estimate the bolometric luminosity from $L_{\rm 12{\mu}m}$ \citep{ich17a}.
For the estimates of the Eddington luminosity ratio $\lambda_{\rm Edd}=L_{\rm bol}/L_{\rm Edd}$,
 we apply the black hole mass of $M_{\rm BH}=6.7 \times 10^8$~$M_{\odot}$, which leads to 
 $L_{\rm Edd}=8.4 \times 10^{46}$~erg~s$^{-1}$.}
}
\end{deluxetable*}

\subsection{X-ray Observations}\label{sec:X-ray}

Although the X-ray observations give us the current AGN activity without the concern of the 
obscuration up to $\log N_{\rm H} \simeq 24$ \citep[e.g.,][]{ric15,ric17}, 
we have not found any previous X-ray observations for Arp~187,
and therefore only the upper bounds obtained from the available all-sky X-ray surveys.

Arp~187 is not in the catalog of \textit{Swift}/BAT 105 month all-sky survey with the limiting flux at the 14--195~keV band of
$f_{\rm 14-195}^{(\rm lim)}=8.0\times 10^{-12}$~erg s$^{-1}$ cm$^{-2}$ \citep{oh18}.
This gives a very shallow upper bound of $L_{14-195}< 2.8 \times 10^{43}$~erg~s$^{-1}$,
which is equivalent to $L_{\rm bol}<2.4 \times 10^{44}$~erg~s$^{-1}$ under the assumption
of $L_{\rm bol}/L_{14-195}=8.47$ \citep{ric17a}.

The \textit{ROSAT} All Sky Survey also shows non detection at the energy range of 0.5--2.0~keV
\citep[$f_{\rm lim}=2.5\times 10^{-12}$~erg s$^{-1}$ cm$^{-2}$;][]{vog99}. This also gives a shallow upper bound
of $L_{\rm 0.5-2}< 8.8 \times 10^{42}$~erg~s$^{-1}$, which is equivalent to 
$L_{\rm bol}<2.8 \times 10^{44}$~erg~s$^{-1}$ using $L_{2-10}/L_{0.5-2}=1.57$
under the assumption of the photon index $\Gamma=1.8$ \citep{ric17} and $L_{\rm bol}/L_{2-10}=20$ \citep{ric17a}.

Other X-ray catalogs such as
 the third \textit{XMM-Newton} serendipitous source catalog \citep[3XMM-DR7;][]{ros16}, 
Chandra Source Catalog \citep[CSC Release 2.0;][]{eva10} do not contain the observations of Arp~187.

\subsection{Black Hole Fundamental Plane}\label{sec:BHfundamental}

The fundamental plane of the black hole gives a relationship among three physical quantities
of $L_{2-10}$, core $L_{\rm 5GHz}$, and the black hole mass $M_{\rm BH}$
\citep[e.g.,][]{mer03,fal04,yua14}.
\cite{ich16} discussed that once the upper-bound of $L_{\rm 5GHz}$ is given, 
we can estimate the upper-bound of $L_{2-10}$ since the black hole mass in Arp~187 is 
estimated to be $M_{\rm BH}=6.7 \times 10^8$~$M_{\odot}$.
The upper bound of $L_{\rm 5GHz} \leq 3.7 \times 10^{37}$~erg~s$^{-1}$
obtained by the VLA observation gives $L_{2-10} \leq 4.0 \times 10^{39}$~erg~s$^{-1}$
using the relation of \cite{yua05}.
This is equivalent to $L_{\rm bol} \leq 8.0 \times 10^{40}$~erg~s$^{-1}$,
indicating that the central engine is already quenched.

\begin{figure*}
\begin{center}
\includegraphics[width=0.7\linewidth]{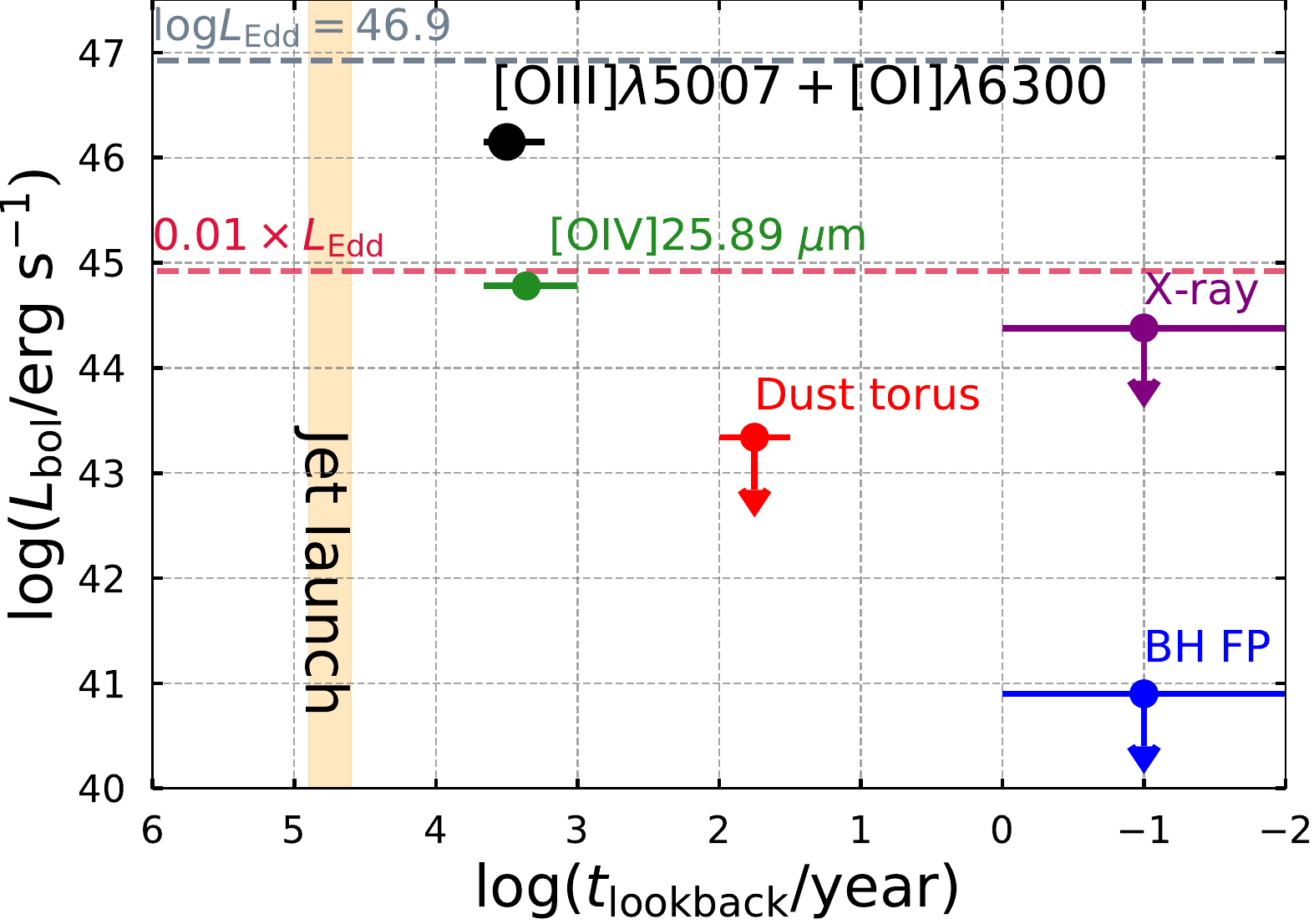}~
\caption{
Longterm light curve of Arp~187 based on the AGN indicators with multi physical scales.
See Table~\ref{tab:AGNindicators} for the details of the each AGN indicator.
}\label{fig:LC}
\end{center}
\end{figure*}

\section{Discussion}\label{sec:discussion}

\subsection{AGN Indicators and Scenarios of Current AGN Activity}

Our multi-wavelength measurements of the AGN indicators show that we have successfully found
the large scale ($>100$~pc) AGN indicators while not for the small scale ones with $\leq 10$~pc.
Table~\ref{tab:AGNindicators} summarizes the AGN indicators for Arp~187.

There are two possible scenarios that can account for these seemingly contradictory results 
between the larger ($>100$~pc) and smaller ($\leq 10$~pc) scales.
 One is that the AGN activity still exists but the emission is highly obscured
  along the line of sight, while being unobscured towards the jet and/or the NLR direction
  \cite[e.g.,][]{sar18a}.
 The other is that the AGN activity in Arp~187 has already been quenched,
 while the narrow line regions still remain bright due to the light-travel time
 from the central engine.
 The first scenario is unlikely for Arp~187 because of the two reasons.
 One is the absence of the AGN torus emission in the \textit{Spitzer}/IRS spectra as well as the other torus indicators　obtained from \textit{WISE}. 
 If the central engine is highly obscured, then the most of the
 emission is absorbed by the dust, and then it re-emits and produces the bump in the mid-IR \citep[e.g.,][]{ich14}, which we could not find.
Another point is the absence of the jet core, which is irrelevant to the concern of absorption and its existence is the ubiquitous trend for AGN \citep{bla79,had11}
while there are some rare exceptions \citep{cor87,dwa09}.
Thus, our results prefer the scenario of dying AGN.

One might argue that the absence of the big blue bump and the
broad emission line region (BLR) is
 due to the absorption by the dusty torus \citep[e.g.,][]{ant93,nag02}.
 It is true in general argument, but since the torus emission is already weak for Arp~187
 as discussed in Section~\ref{sec:Spitzer}, the most plausible explanation is that
 the central engine including the accretion disk and the BLR becomes 
 already very faint or might be diminished.
  
The disappearing timescale of the BLR is poorly known at current stage.
One possible implication of this comes from the observations of changing-look AGN,
which show the AGN type change in human timescale \citep[e.g.,][]{toh76,ant83,all85,lam15,rua16,mac16,yan17}.
One of the findings of the $\sim10$~years long monitoring of changing-look AGN is that
the BLR can disappear within a few years, while
the physical mechanism of the disappearance is still under debate \citep{law18}. 
Thus, in this paper we compile the timescale of BLR as 1--10~years in Table~\ref{tab:AGNindicators}.

\begin{deluxetable*}{cccc}
\tablecaption{Observational Properties of dying AGN, 
fading AGN, and changing-look quasars\label{tab:comparison}}
\tabletypesize{\footnotesize}
\tablewidth{0pt}
\tablehead{
& dying AGN (Arp~187) & fading AGN & changing-look quasar
}
\startdata
  & AGN whose current (small scale)  & AGN with weaker current $L_{\rm AGN}$ & quasars with broad Balmer line  \tablebreak 
 Definition &AGN signature  is dead,  but   &  compared to that of & (dis)appearance associated with\tablebreak
 & past AGN signature is still alive. & past AGN signatures. &  continuum change by a factor of $\sim 10$.
 \\\hline
Small scale ($<10$~pc) AGN signs & dead & alive (but weak) & alive \\
Large scale ($>100$~pc) AGN signs & alive & alive & alive\\
Jet core features & no & yes (?)$^{(\rm B1)}$ & \nodata \\
Jet lobe features & yes & \nodata & \nodata \\
$\Delta L_{\rm AGN}^{\rm (a)}$& $10^{3-5}$ & $10^{1-4}$ & $\sim 10$\\
$\Delta t^{\rm (b)}$ & $\leq 10^4$~yr & $10^{4-5}$~yr (using EELR$^{(1)}$) & $\sim 1$--$10$~yr$^{\rm (C1,C3,C4,C6)}$\\
$L_{\rm AGN}$ (current) & $< 10^{42}$~erg~s$^{-1}$ &　$10^{43-45}$~erg~s$^{-1}$　&
$> 10^{43}$~erg~s$^{-1}$\\
$L_{\rm AGN}$ (before fading) & $\sim 10^{46}$~erg~s$^{-1}$ & $10^{44-46}$~erg~s$^{-1}$ &
$>10^{44}$~erg~s$^{-1}$ \\
Redshift ($z$) & 0.04 & $0.01$--$0.3^{\rm (B5, B10)}$ & $0.01$--$1.0^{\rm (C1,C3,C4,C6)}$\\
\multirow{2}{*}{Origin of variability} & \multirow{2}{*}{viscous} & 
\multirow{2}{*}{viscous} & unknown (thermal?$^{\rm (C8-C10)}$ or\tablebreak
&&&  magnetically elevated disk?$^{\rm (C7)}$)\\ 
Host galaxies& merger remnant & merger system/remnants (?)$^{\rm (B5)}$ & \nodata \\
Number of sources found & \nodata & $\sim20$ sources$^{\rm (B1-B13)}$ & $>40$ sources$^{\rm (C1-C6)}$\\
\multirow{2}{*}{Most known object names} & Arp~187$^{(\rm A1)}$ & IC~2497 (Hanny's Voorwep)$^{\rm (B1-B8)}$ & SDSS~J0159+0033$^{\rm (C1)}$ \tablebreak 
& (NGC 7252?$^{\rm (A2)}$) & Teacup galaxy$^{\rm (B5,B6,B12)}$, etc. & SDSS~J1011+5442$^{\rm (C2)}$, etc. \\
\enddata
\tablecomments{
(1) EELR stands for extended AGN photo-ionized emission line region
with the physical scale of $\sim10$~kpc.
(a) observed or inferred AGN luminosity decline
(b) the timescale of the AGN luminosity decline of $\Delta L_{\rm AGN}$.
``$\cdots$'' in the column represents
that there are no clear observations or consensus from the literature.
}
\tablerefs{
dying AGN: 
(A1) \cite{ich16}, (A2) \cite{sch13b}; 
fading AGN: 
(B1) \cite{joz09},
(B2) \cite{lin09}, 
(B3) \cite{sch10}, 
(B4--B6) \cite{kee12,kee15,kee17}, 
(B7,B8) \cite{sar18a,sar18b}, 
(B9,B10) \cite{sch13,sch16},
(B11) \cite{kaw17},
(B12) \cite{vil18}, 
(B13) \cite{wyl18}; 
changing-look quasar:
(C1) \cite{lam15}, (C2) \cite{run16}, (C3) \cite{rua16}, 
(C4) \cite{mac16}, (C5) \cite{gez17}, (C6) \cite{yan17}, 
(C7) \cite{dex18}, (C8) \cite{nod18}, (C9) \cite{ros18}, (C10) \cite{ste18}
}
\end{deluxetable*}

\subsection{Luminosity Decline and Corresponding Timescales}\label{sec:timescale}

One of our goal is to constrain how rapidly the AGN has faded in Arp~187.
We summarize the long-term luminosity decline of Arp~187 in Figure~\ref{fig:LC}.
This figure shows that the AGN has experienced the drastic luminosity decline 
by the factor of $10^{4}$ times
within $\simeq 10^{4}$~years.

One question arises as to how to describe the sudden quenching of AGN within $\simeq 10^{4}$~years and how this timescale is connected to
the physical properties of the accretion disk of AGN.
We here consider three possible timescales: the orbital, thermal, and the
viscous (inflow) timescales, by following the discussions of \cite{cze06} and \cite{ste18}.
If the accretion flow is roughly Keplerian, the dynamical
timescale of the accretion disk is given by $t_{\rm dyn} \sim 1/\Omega_{\rm K}$,
where $\Omega_{\rm K}$ is the Keplerian orbital angular frequency.
The thermal time scale, which is corresponding of the disk cooling,
is given by $t_{\rm th} \sim 1/\alpha \Omega_{\rm K}$, where $\alpha$
is the viscosity parameter \citep{sha73,kat08}.
The viscous timescale is also given by
$t_{\rm vis} \sim (h/R)^{-2} t_{\rm th}$,
where  $h/R$ is the disk aspect ratio.
The numerical simulations derive the estimation of $\alpha \sim 0.03$
\citep{hir09,dav10},
therefore we use $\alpha_{0.03}=\alpha/0.03$ as the fiducial value.
The disk aspect ratio is typically very small and 
we assume $h/R \sim 0.05$ by following \cite{ste18}.

We first determine the boundary accretion disk radius $R$, 
within which it emits the UV radiation,  
where UV represents the wavelength of $\lambda < 3500$~\AA,
and the typical wavelength range of the big blue bump 
\citep[e.g.,][]{mal82,van01}.
Using the obtained parameter for Arp~187, 
the BH is estimated to be $M_{\rm BH}=6.7 \times 10^8$~$M_{\odot}$
and the maximum Eddington ratio in this study is $\lambda_{\rm Edd}= 0.17$
obtained from the NLR.
The gravitational radius is given by $R_{\rm g}=2GM_{\rm BH}/c^2 \sim 13$~AU.
the AGN bolometric luminosity is governed by $L_{\rm bol}=\eta \dot{M} c^2$,
where $\eta$ is the radiation efficiency.
The typical value is estimated as $\eta \sim 0.1$ \citep{sol82}.
The mass accretion rate $\dot{M}$ is therefore given by 
\begin{equation}\label{Eq:dotM}
\dot{M} \approx 2.8 M_{\odot} {\rm yr}^{-1} \eta_{0.1}^{-1}
M_{\rm BH,0} \lambda_{\rm Edd,0.17}.
\end{equation}
here $\eta_{0.1}=\eta/0.1$,
$M_{\rm BH,0}=M_{\rm BH}/6.7\times 10^{8}M_{\odot}$,
and $\lambda_{\rm Edd,0.17}=\lambda_{\rm Edd}/0.17$.

Assuming a standard thin-disk AGN model,
the disk radius $R$ is linked to the disk temperature $T$ in Equation of (3.57)
of \cite{kat08} written by
\begin{equation}\label{Eq:RvsT}
R = \frac{3GM\dot{M}}{8\pi \sigma T^4},
\end{equation}
where $\sigma$ is a
is a Stefan--Boltzmann constant. 
Combining Equation~(\ref{Eq:RvsT}) and (\ref{Eq:dotM}) with Wien's law 
($(\lambda/{\mbox{\AA}}) = 2.9 \times 10^7/T$), the typical radius $R$ can be given by
\begin{align}
\frac{R}{R_{\rm g}} = 100 \eta_{0.1}^{-1/3} M_{\rm BH,0}^{-1/3}
\lambda_{\rm Edd,0.17}^{1/3}
\lambda_{3500}^{4/3}
\end{align}
where $\lambda_{3500}=\lambda/3500 \mbox{\AA}$.
Thus we apply $R_{100}=R/100R_{\rm g}$ as a fiducial value
for the UV emitting disk size.

We can then parametrize the disk timescales as
 \begin{align}
t_{\rm dyn} &\sim 100~{\rm day}\times M_{\rm BH, 0} R_{100}^{3/2} \\
t_{\rm th} &\sim 10~{\rm yr}\times \alpha_{0.03}^{-1}M_{\rm BH, 0} R_{100}^{3/2} \\
t_{\rm vis} &\sim 1.6\times10^4~{\rm yr} \times \left(\frac{h/R}{0.05}\right)^{-2}
\alpha_{0.03}^{-1}M_{\rm BH, 0} R_{100}^{3/2} 
 \end{align}

 Since Arp~187 does not show the big blue bump and the torus emission
 (whose source of nutrition is the UV photons) anymore at least the
 last a few 10~years (see also Figure~\ref{fig:LC}), 
 the dynamical and the thermal
 time scale are unlikely and too short to be happening.
On the other hand, the viscous timescale seems 
 to be a little longer compared to the quenching timescale of $<10^4$~years.
However, since the NLR traces higher energy UV photons with $>10$~eV,
 the corresponding disk region becomes more inner with $R \lesssim 10$~$R_{\rm g}$, whose viscous timescale is $\lesssim 500$~yrs.
This is consistent with our strongest upper-bound of $\sim3000$~yrs (see Table~\ref{tab:AGNindicators}).
Thus, our rough estimation suggests that the viscous timescale 
most closely matches the observed quenching timescale.

\subsection{Comparison between Dying AGN and Fading AGN}

The absence of small scale AGN signatures and the
timescale discussed in Section~\ref{sec:timescale}
support that Arp~187 has been in a later fading phase,
or ``dying'' phase compared to other fading AGN since the UV emitting
region in Arp~187 likely disappeared, and the large scale AGN indicators
are observable as remnant signs of AGN. On the other hand,
the previously reported fading AGN are considered to be
earlier fading stage since they still host 
either of clear UV, MIR, or X-ray emission in the core, suggesting that 
the UV emitting region is still alive \citep[e.g.,][]{kee15,kee17,sar18a}.
We have summarized the comparison of 
properties of our dying AGN and fading AGN in Table~\ref{tab:comparison}.

One important question is how many　
such ``dying'' AGN have been already 
reported in the  population of fading AGN from the literature.
While most of fading AGN still show MIR or X-ray emission in the core,
one fading AGN in NGC~7252 might fulfill ``dying'' AGN criterion.
\cite{sch13b} showed that NGC~7252 hosts large [OIII]$\lambda5007$
bright nebulae which belong to a stream of tidal-tail gas falling back 
to the host galaxy. The bright [OIII] nebulae require the AGN luminosity
larger than $L_{\rm bol}>5 \times 10^{42}$~erg~s$^{-1}$ while
the current X-ray upper-bound gives $L_{\rm bol}<5 \times 10^{40}$~erg~s$^{-1}$.
Considering the nebulae distance, NGC~7252 might have experienced
a luminosity decline by two orders of magnitude over the past $10^{4-5}$~years,
and the current AGN activity is well below $L_{\rm bol}<10^{42}$~erg~s$^{-1}$,
suggesting that the central engine is already dead.

Note that the past inferred AGN luminosity
is completely different between Arp~187 and NGC~7252.
Arp~187 reached a quasar-like luminosity with $L_{\rm bol}>10^{46}$~erg~s$^{-1}$, 
whereas that of NGC~7252 is well below the quasar level or more likely
Seyfert level luminosity with $L_{\rm bol}>5 \times 10^{42}$~erg~s$^{-1}$.
\cite{sch13b} also reported that the central gas disk of NGC~7252 
contains the large amount of molecular gas with $>10^9$~$M_{\odot}$,
suggesting that the AGN feedback activity have failed to remove the gas
in the host galaxy. Therefore the gas content difference between 
NGC~7252 and Arp~187 would be a good testbed to investigate the effect of AGN feedback since these two galaxies have already experienced the one-cycle of AGN 
activity for at least $10^{5}$~years \citep{sch15} but with different AGN luminosity.

\subsection{Comparison of the Causes of Luminosity Changes
between Dying AGN and Changing-Look Quasars}

It is worthwhile to note the difference of the 
accretion mechanism between our dying AGN in Arp~187
and a recently discovered class of ``changing-look quasars''
in which the strong UV bump and broad emission lines
associated with optically bright quasars either appear or disappear on timescales of years \citep[e.g.,][]{lam15,mac16,gez17,yan17}.
The physical processes causing these changing-look phenomena
are hotly debated, but the physical changes in the accretion disk
is the likely cause rather than changes in obscuration
\citep[e.g.,][]{dex18, law18}.

\cite{ste18} recently discovered one changing-look quasar
WISE J1052+1519, and carefully discussed the possible
disk timescales matching the year-timescale.
They found that the dynamical timescale
is several weeks, which is therefore too shorter, 
while the viscous timescale, which would be responsible
for the luminosity change of dying AGN, is far too long.
Instead, a few year long thermal timescale would
be a plausible one matching the observed year scale variability.
The similar origin is also proposed for a 
different changing-look AGN Mrk~1018 \citep{nod18}.
Therefore, both dying and changing-look AGN show
the luminosity change, but their luminosity changes are likely
based on the different physical mechanisms of the accretion disk.
Those property differences between dying AGN and changing-look quasars
are also summarized in Table~\ref{tab:comparison}.

\subsection{Comparison between Dying AGN and Remnant Radio Sources}
Our study suggests that the absence of radio core at the center of galaxies
would be a good indicator for searching for a dying AGN.
While they are very rare, some authors have already found candidates
of radio galaxies without clear radio core signs, or so called remnant radio sources.
\cite{cor87} showed that IC~2476 has double radio lobes
with a separation of 560~kpc, but without a
clear radio core at the location of the host galaxy.
In addition, \cite{dwa09} have conducted a search for remnant radio sources
using VLA 74~MHz survey (VLSS; 80~arcsec spatial resolution) 
and NRAO VLA Sky Survey (NVSS; 45~arcsec spatial resolution)
through the search of very steep radio sources whose
spectral index is $\alpha < -1.8$ (where
$f_{\nu} \propto \nu^{\alpha}$) between 74~MHz and 1.4~GHz.
Out of the $\sim 10^4$ parent sample, they found 10 such candidates
and the spectral age estimation of jet lobes gives the fading age of $>10$~Myr.

The timescale found in remnant radio sources is
at least two orders of magnitude longer than the fading phase traced
for Arp~187.
This discrepancy is natural because their steep spectral selection method is 
 sensitive to longer jet age with $>10^7$~years
at the frequency of $\sim 100$~MHz \citep[e.g.,][]{jam08}
and their moderate spectral resolution of 80~arcsec is equivalent to
the physical angular size of $\sim260$~kpc, or the corresponding kinematic
age of $t_{\rm dyn} \sim 8 \times 10^6$~years at $z \sim 0.2$, where
most of the sample are found.
Thus, previous radio studies are sensitive to trace
much longer AGN activities with $>10$~Myr.
Considering the typical quasar lifetime of $\sim30$~Myr \citep[e.g.,][]{
hop06,ina18}, remnant radio sources are more suitable to
trace a comparable timescale of the 
AGN lifetime rather than the AGN fading 
timescale traced in this study.

Recently, higher spatial resolution
surveys are ongoing using  Low-Frequency Array \citep[LOFAR;][]{van13}.
LOFAR covers the largely unexplored low-frequency range between 10--240~MHz,
and has the resolution of $\sim6$~arcsec at 150~MHz.
This would give an opportunity to search for smaller sized, which is equivalent to younger,
dying radio sources at the age of $\sim1$~Myr, and the initial-stage surveys have already found new remnant radio source candidates \citep[e.g.,][]{mah18}
and future LOFAR surveys would give us a 
more statistically significant number
of such younger remnant radio sources and they would help to
create a more complete picture of both of AGN lifetime and dying phase
\citep[also see a recent review by][]{mor17}.

\subsection{Future observations}
There is some room for the further constraints of the AGN activities of Arp~187.
\textit{JWST}/MIRI will give us the nuclear MIR spectra with the least host galaxy contamination
with the great sensitivity. The X-ray satellite \textit{NuSTAR} is going to constrain the
current AGN activity.
  Recently, thanks to the great sensitivity at $E>10$~keV,
   \textit{NuSTAR} revealed that a fading AGN candidate in IC~2497 
  is actually a Compton-thick AGN \citep{sar18}, 
  whose nuclear X-ray emission could not be discovered in the previous X-ray
 satellites such as \textit{Chandra}, \textit{XMM/Newton}, and even with \textit{Suzaku} \citep{sch10}.
Although the same discovery is unlikely for Arp~187 since the MIR emission is known to be
considerably weak, \textit{NuSTAR} will constrain the nuclear activity down to 
 $\log (L_{2-10}/{\rm erg}~{\rm s}^{-1}) \simeq 42$ even with the Compton-thick absorption of 
$\log N_{\rm H} \simeq 24.3$. Finally, the optical or near-IR IFU will also give us the detailed NLR size
which is poorly constrained with the current study.

\section{Conclusion}

We have compiled the multi-wavelength AGN signatures of dying AGN candidate Arp~187,
based on the combinations of the newly conducted ALMA observations
 as well as the archival VLA 5--10~GHz data,
6dF optical spectrum, \textsc{Neowise} and \textsc{Allwise} IR data.
Our results show that the AGN in Arp~187 is a bona fide dying AGN,
whose central engine is already dead,
 but the large scale AGN indicators are still observable as the remnant 
 of the past AGN activity.
The central engine of Arp~187 has experienced the drastic luminosity decline
by a factor of $10^{3-5}$ within the last $10^{4}$~years.
Our rough estimation suggests that the viscous timescale
most closely matches the obtained timescale in this study.
This supports that Arp~187 has been in more later 
fading phase whose UV emitting region in the accretion disk
is likely to be almost disappeared, while other fading AGN show clear signs
that UV emitting region is still alive.

\acknowledgments

We thank the anonymous referee for helpful suggestions that strengthened the paper.
We thank Kazunori Akiyama, Luca Comisso, Kohei Inayoshi, Masatoshi Imanishi, 
Mitsuru Kokubo, Hiroshi Nagai, and Hirofumi Noda for fruitful discussions.
We also would like to thank Jaejin Shin for giving us the spectral fitting result of the optical spectrum.

This paper makes use of the following ALMA data: ADS/JAO.ALMA\#2015.1.01005.S. 
ALMA is a partnership of ESO (representing its member states), NSF (USA) 
and NINS (Japan), together with NRC (Canada) and NSC and ASIAA (Taiwan) 
and KASI (Republic of Korea), in cooperation with the Republic of Chile.  
The Joint ALMA Observatory is operated by ESO, AUI/NRAO and NAOJ.
In addition, this paper includes observational material taken with VLA, 
which is one of the NRAO instruments.
The National Radio Astronomy Observatory is a facility 
of the National Science Foundation operated 
under cooperative agreement by Associated Universities, Inc.
This paper also makes use of data products from the \textit{WISE}, 
which is a joint project of the University of California, Los Angeles, 
and the Jet Propulsion Laboratory/California Institute of Technology, 
funded by the National Aeronautics and Space Administration.

This work is supported by Program for Establishing a Consortium
for the Development of Human Resources in Science
and Technology, Japan Science and Technology Agency (JST) and
 is partially supported by Japan Society for the Promotion of Science (JSPS) KAKENHI (18K13584; KI, 18J01050; YT, 16K17672; MS). 
 K.M. is supported by JSPS Overseas Research Fellowships.

%

\vspace{5mm}
\facilities{ALMA, VLA, \textit{WISE}, \textit{Spitzer}, \textit{Chandra}, \textit{XMM-Newton}, \textit{Swift}/BAT}


\software{astropy \citep{ast13}, Matplotlib \citep{hun07}, Pandas \citep{mck10}
          }

\bibliographystyle{aasjournal}
\bibliography{arp187}



\end{document}